\documentclass[aps,prl,twocolumn,groupedaddress,showpacs]{revtex4}

\usepackage{graphicx}

\begin{document}

\title{Neutral Plasma Oscillations at Zero Temperature}

 \author{S. D. Bergeson}
 \author{R. L. Spencer}
 \affiliation{Department of
 Physics and Astronomy, Brigham Young University, Provo, UT 84602}

\date{\today}

\begin{abstract}
We use cold plasma theory to calculate the response of an
ultracold neutral plasma to an applied rf field. The free
oscillation of the system has a continuous spectrum and an
associated damped quasimode.  We show that this quasimode
dominates the driven response. We use this model to simulate
plasma oscillations in an expanding ultracold neutral plasma,
providing insights into the assumptions used to interpret
experimental data [Phys. Rev. Lett. {\bf 85}, 318 (2000)].
\end{abstract}


\maketitle

Recent experimental
\cite{rolston98,killian99,kulin00,robinson00,killian01,dutta01}
and theoretical work
\cite{cote00,greene00,tkachev01,hahn01,robicheaux02,mazevet02,kuzmin02}
has studied the formation and evolution of ultracold plasmas.  In
the laboratory, cold plasmas are created either by directly
photo-ionizing laser-cooled atoms, or by exciting the atoms to
high-lying Rydberg states that spontaneously ionize. A fraction
of the electrons escape the plasma and the resulting electric
field drives the ion expansion.  Models of the expansion suggest
that the density profile in the plasma is approximately Gaussian,
and that it expands in a self-similar manner. It can be expressed
as
  \begin{equation}
  \label{eqn:expansion} n(r,t)= {{n_0 \sigma_0^3}\over{\sigma^3}}
  \exp[-r^2/2\sigma^2] ~ ,
  \end{equation}
\noindent where $\sigma=\sqrt{\sigma_0^2 + v^2t^2}$ is the
time-dependent width of the distribution, $v$ is the asymptotic
expansion velocity, and $t$ is time.

The experimental justification of this density profile for the
case of an expanding plasma is based on the plasma's response to
a spatially uniform applied rf field. In those experiments, Xe
atoms initially cooled to $\sim 10 \mu$K were ionized by a dye
laser. The initial electron energy ($E_e/k_B$) ranged from a few
to 1000 K, and the initial electron density ranged from $0.2$ to
$2.5 \times 10^{9}$ cm$^{-3}$.  The plasmas were nearly
charge-neutral, and at electron energies above 70 K, the kinetic
energy of the electrons allowed significant loss.  The resulting
net positive charge in the cloud drove the plasma expansion.

As the plasma expanded and its density decreased, the applied rf
field pumped energy into the plasma.  The heating was assumed to
be largest where the applied rf frequency matched the local plasma
frequency.  Because of collisions in the plasma, the local
heating presumably raised the overall plasma temperature, and a
small number of the more weakly-bound electrons were ejected. The
experiment measured the rate at which electrons were ejected from
the expanding plasma as a function of time for a fixed applied rf
frequency.  This signal was presumed to be proportional to the
rate at which the rf field heats the plasma, and was a measure of
the weighted time-dependent density profile.

The peak of this signal was interpreted to correspond to the
``average'' density of the plasma,
  \begin{equation}
  \bar{n}(t) = 2^{-3/2}n(0,t) =
  {{n_0\:\sigma_0^3}
  \over{\left[2\left(\sigma_0^2+v^2t^2\right)\right]^{3/2}}}
  \label{eqn:nbar}
  \end{equation}
with $v$ a constant.  This density was experimentally determined
by setting the applied frequency $\omega$ equal to the average
plasma frequency, $\omega_p$, and using the relation
$\bar{\omega}_p = \sqrt{q^2 \bar{n}(t) / m_e \epsilon_0}$, where
$q$ is the electron charge, $m_e$ is the electron mass, and
$\epsilon_0$ is the permittivity of free space.  The derived
density, $\bar{n}$, is time dependent.  For a different applied
rf frequency $\omega$, the signal peaks at a different time. A
least-squares fit of $\bar{n}(t)$ to Eq. (\ref{eqn:nbar}) gives
the initial density $n_0$ and the expansion velocity $v$.

In this paper, we examine the validity of the assumptions used in
this interpretation by presenting the response of the system
predicted by cold plasma theory.  This theory is good for plasmas
in which $(\lambda_D/\sigma)^2 \ll 1$, where $\lambda_D$ is the
Debye length. In the experiments we are simulating,
$(\lambda_D/\sigma)^2 \approx 10^{-3} - 10^{-1}$, making cold
plasma theory a good way to model the collective response of the
system.

We consider spherically symmetric charge-neutral Gaussian
distributions of cold ions and electrons to which an rf-electric
field in the $z$-direction is applied. The ions are taken as
fixed on the time-scale of the rf oscillation. The fluid
equations describing the electrons are
\begin{eqnarray}
 {{\partial n}\over{\partial t}} + \nabla \cdot
 \left(n\vec{v}\right)=0 \label{eqn:f1} \\
 \nabla^2 \phi = -q \left(n - n_i \right) / \epsilon_0 \label{eqn:f2} \\
 m \left( {{\partial \vec{v}}\over{\partial t}} + \vec{v}\cdot
 \nabla \vec{v}\right) = q\vec{E}-\gamma m \vec{v}
 \label{eqn:f3}.
 \end{eqnarray}
\noindent In these equations $n=n(r,t)$ is the electron density,
$n_i=n_i(r)$ is the ion density (assumed to be fixed in time),
$\vec{v}=\vec{v}(\vec{r},t)$ is the electron velocity,
$\vec{E}=\vec{E}(\vec{r},t)$ is the total electric field (the sum
of the external field and that generated by the plasma response),
and $\gamma$ is a phenomenological damping rate. Some effects of
finite temperature, such as electron-electron collisions and
electron-ion collisions, are approximated by the the parameter
$\gamma$.

In the experiment by Kulin et al. \cite{kulin00}, the applied
$\vec{E}$ was uniform in space and oscillating in the
$\hat{z}$-direction. Assuming that this applied field is small,
the fluid equations can be linearized by assuming that the
density and the potential are of the form
$n(r,\theta,t)=n^{(0)}(r) + \delta n(r) \cos{\theta}e^{-i \omega
t}$, where $n^{(0)}(r)$ is the equilibrium electron density, and
$\phi(r,\theta,t)=\delta\phi(r) \cos{\theta}e^{-i \omega t}$
(spherical harmonics with $\ell=1$). After some algebra to
linearize the fluid Equations (\ref{eqn:f1})-(\ref{eqn:f3}), the
potential produced by the electrons is given by the expression
  \begin{equation}
  {\cal{L}} \delta \phi - {{ \left[{{d}\over{dr}}\omega_p^2(r)\right]
  {{\partial \delta \phi}\over{\partial r}} }\over{\omega^2 -
  \omega_p^2(r) + i \omega \gamma}} = - {{
  \left[{{d}\over{dr}}\omega_p^2(r)\right] E_0
  }\over{\omega^2 - \omega_p^2(r) + i \omega \gamma}}
  \label{eqn:phi}
  \end{equation}
\noindent where the operator $\cal{L}$ is defined as
  \begin{equation}
    {{d^2}\over{dr^2}} + {{2}\over{r}}{{d}\over{dr}} -
    {{2}\over{r^2}}
  \end{equation}
\noindent and where $\omega_p(r) = \sqrt{q^2 n^{(0)}(r)/m_e
\epsilon_0}$ is the plasma frequency. The boundary conditions are
that $\delta\phi(0)=0$ and that at infinity the field is that of a
dipole, $\delta\phi(r) \propto 1/r^2$.

\begin{figure}
\includegraphics[angle=270,width=3in]{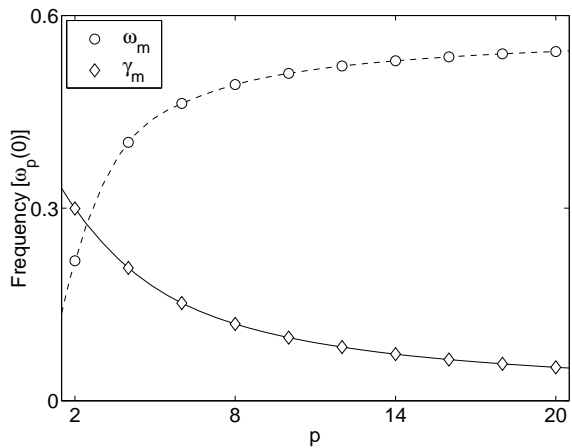}
\caption{\label{fig:pplot} Quasimode frequency, $\omega_m$
($\circ$), and damping rate, $\gamma_m$ ($\diamond$), for density
profiles given by Eq. (\ref{eqn:pprofile}).  For $p=2$, the
distribution is Gaussian.  As $p$ increases, the distribution
approaches a ``top-hat,'' the quasimode frequency increases, and
the quasimode damping decreases (see text).}
\end{figure}

Because this is a driven system, we first look for normal modes.
Notice, however, that with $E_0=0$ and $\gamma=0$ (undamped free
response) this mode equation has a continuous spectrum, similar
to the diocotron mode equation in non-neutral
plasmas\cite{schecter99a,spencer97,corngold95,briggs70,case60,kelvin80}.
And, as with the diocotron mode, this equation also has a damped
quasimode. Following Ref. \cite{spencer97}, Eq. (\ref{eqn:phi})
can be solved along a contour in the complex $r$-plane to uncover
this damped quasimode. Figure \ref{fig:pplot} shows the frequency
and damping rate of this quasimode for a sequence of density
profiles of the form
  \begin{equation}
  n(r) = n_0 e^{-r^p/2 \sigma^p}
  \label{eqn:pprofile}
  \end{equation}
The real and imaginary parts of the quasimode frequency are well
approximated by the simple formulas
$\omega_m/\omega_p(0)=1/\sqrt{3}-0.668/p-0.102/p^2$ and
$\gamma_m/\omega_p(0)=1.077/p-0.959/p^2$.

Note that the Gaussian ($p=2$) quasimode is heavily damped and
that as $p$ becomes large (making the density profile approach a
step-profile) the damping rate goes to zero and the frequency
approaches $\omega_m=\omega_p(0)/\sqrt{3}$.

\begin{figure}
\includegraphics[angle=270,width=3in]{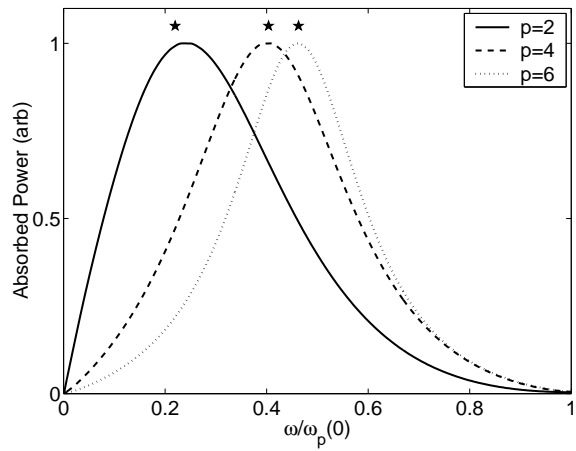}
\caption{\label{fig:mode}Power absorbed by the (non-expanding)
plasma as a function applied frequency.  The solid line shows the
absorbed power for a Gaussian density profile, $p=2$ in Eq.
(\ref{eqn:pprofile}). Also shown is the absorbed power for
super-Gaussian profiles, with $p=4,6$. The mode damping decreases
with increasing $p$. These plots are generated with
$\gamma/\omega_p^{(0)}=0.01$.  For larger values of $\gamma$,
these curves are somewhat broader and shifted to slightly higher
$\omega$.}
\end{figure}

The quasimode dominates the free response of a system with a
continuous spectrum \cite{schecter00,schecter99}. It also strongly
influences the driven response, as can be seen in Fig.
\ref{fig:mode}.  This figure shows the rate at which a plasma
absorbs power from an applied rf field as a function of the
frequency of that field. To calculate this power, we note that
the total electric field (the applied field plus the plasma
response) is given by
  \begin{equation}
  \vec{E}(r,t)=\left(\hat{z}E_0 \cos\theta - \nabla\delta\phi\right)
  e^{-i\omega t}. \label{eqn:field} \end{equation}
\noindent By using the linearized Eq. (\ref{eqn:f3}) to compute
the current density $\vec{J}=qn\vec{v}$ in the plasma, the power
density can be written as
  \begin{equation} \left<P\right> =
  {{1}\over{2}} Re \left( \vec{E}\cdot\vec{J}^{*} \right) =
  {{\epsilon_0}\over{2}} \left|\vec{E}\right|^2
  {{\omega_p^2(r)\gamma}\over{\omega^2+\gamma^2}}.
  \label{eqn:powdens}
  \end{equation}
The influence of the quasimode on the absorbed power can be seen
by integrating this power density over all space for various
values of the driving frequency $\omega$. Although strongly
damped, the quasimode still influences how the plasma responds to
the rf field.  Figure \ref{fig:mode} shows the calculated power
absorbed by the plasma as a function of the applied rf frequency
for a non-expanding plasma. In each case the peak of the absorbed
power occurs near the frequency of the quasimode. Notice that
these resonances occur at quite small values of the density,
rather than in the body of the density distribution as assumed by
Kulin, {\it et al}.  The experiment used $\omega/\omega_p(0)=0.6$

We now use these cold driven solutions of Eq. (\ref{eqn:phi}) to
simulate the plasma expansion experiment.  In order to do this,
we need an expression for the electron density, $n(r,t)$ and the
phenomenological damping rate $\gamma$. For both of these we use
the model of Robicheaux and Hanson \cite{robicheaux02}. They
simulated an expanding ultracold neutral plasma to study
three-body recombination and electron heating.  An important
result of their paper is that the electron temperature scales
with density as $T_e \propto n^{-1/3}$, which makes the Coulomb
logarithm ($\ln\Lambda$) constant. We use this result in our
simulation of the collective response of the plasma to an applied
field.  It is likely that the phenomenological damping rate,
$\gamma$, is proportional to the electron collision rate, $\nu$.
The scaled collision rate in the plasma is $\nu/\omega_p \propto
\sqrt{n/T^3}\ln\Lambda$, which is constant in this temperature
scaling.  For typical experimental parameters, this gives
$\gamma/\omega_p(0) \approx 1/100$.  Another important result from
the work of Robicheaux and Hanson is that most of the ions in the
plasma experience an electric field linearly proportional to
$r$.  Given the known initial conditions in the plasma, this
leads directly to the result that the plasma density is Gaussian,
and that it expands in a self-similar manner described by Eq.
(\ref{eqn:expansion}) \cite{note01}.

\begin{figure}
\includegraphics[angle=270,width=3in]{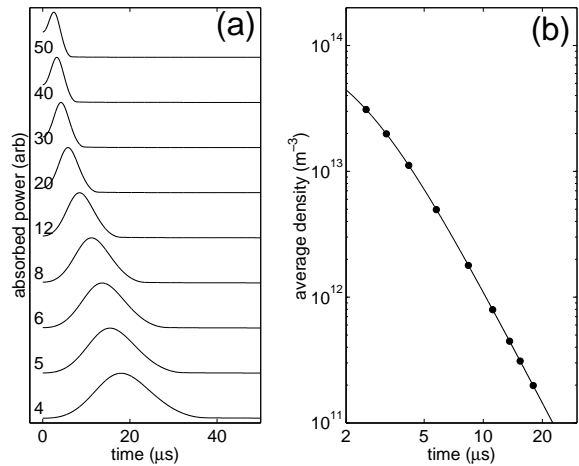}
\caption{(a) Calculated plasma heating rate.  The applied
frequencies $\omega/2\pi$ in MHz are shown to the left of each
trace. The traces are offset vertically for clarity.  The plasma
conditions are $n_0=10^{15}$ m$^{-3}$, $v=100$ m/s,
$\sigma_0=220\;\mu$m, and $\gamma/\omega_p(0)=0.01$.  (b) Average
plasma density as a function of time.  The circles ($\bullet$) are
taken from the data in (a).  The applied frequency is converted
to a density, $\bar{n}(t)$, with $t$ given by the peak of the
signal (see text). The solid line is a fit of these points to Eq.
(\ref{eqn:nbar}). The fitted density is $\bar{n}=1.1\times
10^{14}$ m$^{-3}$ or $n_0=3.1 \times 10^{14}$ m$^{-3}$, about a
factor of $3$ too low. The fitted expansion velocity is $v=100$
m/s. \label{fig:expansion1}}
\end{figure}

We simulate the plasma expansion experiment as follows.  The
plasma expansion velocity ($v$), the rf frequency ($\omega$) and
rf amplitude ($E_0$) are held constant.  At a particular time
$t$, we insert the density from Eq. (\ref{eqn:expansion}) into
Eq. (\ref{eqn:phi}) to find the potential $\delta\phi$ and the
total electric field [Eq. (\ref{eqn:field})].  We integrate the
power density [Eq. (\ref{eqn:powdens})] over the spatial
coordinates to get the plasma heating rate at that time.  We
increment time and repeat the heating rate calculation to
generate one of the time-sweeps shown in Fig.
\ref{fig:expansion1}(a).  Repeating these calculations for a range
of applied field frequencies generates all of the time sweeps
shown in the figure.

We convert the time sweeps in Fig. \ref{fig:expansion1}(a) into
average density determinations.  At the peak of each time sweep
we set the applied frequency $\omega$ equal to the average plasma
frequency $\bar{\omega}_p = \sqrt{q^2 \bar{n} / m_e \epsilon_0}$
to calculate the average density $\bar{n}$, as was done in the
experiment \cite{kulin00}. With this conversion from frequency to
density, the simulated average density $\bar{n}(t)$ is plotted in
Fig. \ref{fig:expansion1}(b). The data is fit to Eq.
(\ref{eqn:nbar}) using a least-squares method to extract the
initial central density, $n_0$, and expansion velocity $v$.

The shapes of the calculated plasma heating rates shown in Fig.
\ref{fig:expansion1}(a) match the experimental ones from
Reference \cite{kulin00}, and the fit of $\bar{n}(t)$ in Fig.
\ref{fig:expansion1}(b) to Eq. (\ref{eqn:nbar}) is excellent
\cite{note03}. The expansion velocity extracted from the fit
exactly matches the known velocity that we put into the
simulation, independent of the value of the scaled damping rate
$\gamma/\omega_p(0)$. If the Robicheaux and Hanson temperature
scaling is correct, the plasma response to the rf field
accurately reproduces the true asymptotic expansion velocity.
However, different temperature scalings make the fit worse. If
instead of using $T \propto n^{1/3}$ in determining $\gamma$, we
use $T \propto n^{2/3}$, the fitted velocity increases by roughly
a factor of two, depends on $\gamma/\omega_p(0)$, the extracted
$\bar{n}(t)$ does not follow Eq. (\ref{eqn:nbar}), and the
simulation does not fit the experiment.

The initial central density $n_0$ extracted from the fit of
$\bar{n}(t)$ to Eq. (\ref{eqn:nbar}) is quite sensitive to the
scaled damping factor, $\gamma/\omega_p(0)$. It scales
approximately as $e^{2\gamma/\omega_p(0)}$.  It also depends on
the particular temperature scaling used in the simulation, and can
vary by orders of magnitude.  For the Robicheaux and Hanson
temperature scaling and with $\gamma/\omega_p(0) = 0.01$, the
fitted central density underestimates the known central density
in our simulation by a factor of three.

Cold plasma theory only addresses the collective behavior of the
system.  Another effect that might be involved in the resonant
behavior seen in the experiment is that the electron motion in
the plasma resonates with the driving field. Robicheaux and
Hanson estimate that in the bulk of the plasma the potential
satisifes $q\phi /k_B T = r^2/2 \sigma^2$. This means that most
of the particles are in an isotropic harmonic oscillator well
with natural frequency $\omega_0 = \omega_p(0) \lambda_D(0)/
\sigma $, where $\lambda_D(0)$ is the Debye length at the center
of the cloud. If the peaks in the experiment were due to particle
resonance at $\omega=\omega_0$ in the expanding plasma, the
experiment does not indicate a uniform expansion \cite{note02}.

However, in experiments where $\lambda_D(0) \ll \sigma$, such as
in the present simulations, we would expect collective effects
like the quasimode, to dominate over individual particle effects
like a resonance at $\omega=\omega_0$. But even though collective
effects should dominate these plasmas, particle resonance can
play a role, as it does in the case of electron Bernstein wave in
magnetized plasmas \cite{krall73}. Calculating this effect
requires a kinetic theory treatment that is beyond the scope of
this paper.  But we note that even in kinetic theory the particle
resonance will not be sharp because the well is not perfectly
parabolic, causing the orbit frequency to vary with the energy
and angular momentum.  Averaging this resonance over the
distribution function blurs its effect and makes the resonances at
$\omega_0$, $2 \omega_0$, etc. less pronounced.

It would be very helpful to have direct measurements of the ion
density distribution, such as with optical detection. These
measurements could provide benchmark data against which future
simulations can be compared. Also, as mentioned previously, full
kinetic simulations are probably required to address the particle
orbit problem.  Such measurements and simulations are currently
underway in our laboratory.

In conclusion, we have used a simulation based on cold plasma
theory to calculate the response of an ultracold plasma to an
applied rf field.  When $T \propto n^{1/3}$, the simulation
accurately reproduces the experimental data of Kulin et al.
\cite{kulin00}, providing an important consistency check of the
model of Robicheaux and Hanson \cite{robicheaux02}.

The model confirms that the plasma response accurately reproduces
the asymptotic velocity of an expanding plasma, and determines
the initial plasma density to within a factor of two or three.
The plasma response is dominated by the quasi-mode.  Even though
the mode is strongly damped, it shifts the plasma resonance to
lower densities than otherwise expected from a local-density
approximation.

This work is supported in part by grants from the Research
Corporation and from the National Science Foundation under Grant
No. PHY-9985027.


\begin{thebibliography}{99}
\bibitem{note01}
In Reference \cite{robicheaux02}, the density is written as
$n(r,t)=N_i[\beta(t)/\pi]^{3/2}\times \exp[-\beta(t)r^2]$, where
$N_i$ is the number of ions.  The function $\beta(t)$ is given by
Eq. (2) in that paper, and can be solved analytically.  For the
given boundary conditions, it is $\beta(t)=1/2(\sigma_0^2 +
v^2t^2)$, making their expression for density identical to Eq.
(\ref{eqn:expansion}) in this work.
\bibitem{note03}
The calculated plasma heating rates shown in Fig.
\ref{fig:expansion1}(a) match the experimental rates when
recombination can be neglected ($E_e/k_B T > 70$K
\cite{kulin00,killian01}).  Recombination depletes the
low-temperature part of the electron distribution.  This
translates into experimentally measured heating rates that fall
off more sharply than those shown in Fig.
\ref{fig:expansion1}(a), measured average density functions
$\bar{n}(t)$ that do not match Eq. \ref{eqn:nbar}, and expansion
velocities greater than expected from the initial electron energy.
\bibitem{note02}
In the asymptotic limit where $\sigma \approx t \sqrt{k_B
T_e/M_i}$, the model of Robicheaux and Hanson predicts that the
resonant frequency is given by $\omega_0 = \sqrt{M_i/m_e}/t$. If
the resonances in the experiment were given by this response, the
density profile would have to be either distinctly non-Gaussian,
or experience a complicated expansion in time.
\end{thebibliography}
\end{document}